\documentclass[11pt]{article}

\usepackage{amsmath}
\usepackage{graphicx}
\usepackage{amsfonts}
\usepackage{amssymb}
\usepackage{epsfig}
\usepackage{color}
\usepackage{psfrag}
\usepackage{epstopdf}

\usepackage{caption}
\usepackage{subcaption}

\newcommand{\EQ}{\begin{equation}}
\newcommand{\EN}{\end{equation}}
\newcommand{\be}{\begin{equation}}
\newcommand{\ee}{\end{equation}}
\newcommand{\bea}{\begin{eqnarray}}
\newcommand{\eea}{\end{eqnarray}}

\DeclareMathOperator*{\SumInt}{%
\mathchoice%
  {\ooalign{$\displaystyle\sum$\cr\hidewidth$\displaystyle\int$\hidewidth\cr}}
  {\ooalign{\raisebox{.14\height}{\scalebox{.7}{$\textstyle\sum$}}\cr\hidewidth$\textstyle\int$\hidewidth\cr}}
  {\ooalign{\raisebox{.2\height}{\scalebox{.6}{$\scriptstyle\sum$}}\cr$\scriptstyle\int$\cr}}
  {\ooalign{\raisebox{.2\height}{\scalebox{.6}{$\scriptstyle\sum$}}\cr$\scriptstyle\int$\cr}}
}

\begin{document} \setcounter{page}{0}
\topmargin 0pt
\oddsidemargin 5mm
\renewcommand{\thefootnote}{\arabic{footnote}}
\newpage
\setcounter{page}{0}
\topmargin 0pt
\oddsidemargin 5mm
\renewcommand{\thefootnote}{\arabic{footnote}}
\newpage
\begin{titlepage}
\begin{flushright}
\end{flushright}
\vspace{0.5cm}
\begin{center}
{\large {\bf Persistent oscillations after quantum quenches in $d$ dimensions}}\\
\vspace{1.8cm}
{\large Gesualdo Delfino$^{1,2}$ and Marianna Sorba$^{1,2}$}\\
\vspace{0.5cm}
{\em $^1$SISSA -- Via Bonomea 265, 34136 Trieste, Italy}\\
{\em $^2$INFN sezione di Trieste, 34100 Trieste, Italy}\\
\end{center}
\vspace{1.2cm}

\renewcommand{\thefootnote}{\arabic{footnote}}
\setcounter{footnote}{0}

\begin{abstract}
\noindent
We obtain analytical results for the time evolution of local observables in systems undergoing quantum quenches in $d$ spatial dimensions. For homogeneous systems we show that oscillations undamped in time occur when the state produced by the quench includes single-quasiparticle modes and the observable couples to those modes. In particular, a quench of the transverse field within the ferromagnetic phase of the Ising model produces undamped oscillations of the order parameter when $d>1$. For the more general case in which the quench is performed only in a subregion of the whole $d$-dimensional space occupied by the system, the time evolution occurs inside a light cone spreading away from the boundary of the quenched region as time increases. The additional condition for undamped oscillations is that the volume of the quenched region is extensive in all dimensions. 
\end{abstract}
\end{titlepage}

\newpage

\tableofcontents

\section{Introduction}
A main question for the nonequilibrium dynamics of extended quantum systems is whether time evolution eventually leads to some form of relaxation \cite{BMcD} or can produce a different behavior. The difficulty of the question calls for analytical study in the basic nonequilibrium setting, which corresponds to quantum quenches\footnote{See \cite{SPS,CC} for early applications of this terminology inspired by thermal quenches in classical statistical systems.}. Here an isolated extended system is in the ground state $|0\rangle$ of its Hamiltonian $H_0$ until the time $t=0$, when the instantaneous change of an interaction parameter leads to the new Hamiltonian $H$ that rules the unitary evolution for $t>0$. The procedure dynamically generates the post-quench state and leaves no ambiguities in the initialization of the nonequilibrium evolution. 

The way to analytically study the quench dynamics on general grounds was found in \cite{quench}, where the case of translation invariant one-dimensional systems was considered. The theory is perturbative in the quench size $\lambda$, and crucially incorporates the fact that normally the quasiparticle modes interact (even at equilibrium, namely at $\lambda=0$). In particular, it was found that, under conditions determined by the theory and including quasiparticle interaction, local observables (e.g. the order parameter) exhibit oscillations undamped in time. It was argued in \cite{oscill} that these oscillations appearing at leading order in $\lambda$ are not washed out by higher orders and stay undamped also when $\lambda$ is not small. Undamped oscillations were then numerically observed over hundreds of periods within the time scale accessible to the simulation of a large quench of the longitudinal magnetic field in the Ising chain \cite{Jacopo}, a basic candidate discussed in \cite{quench,DV}. 

In this paper we consider the case of $d$ spatial dimensions. We show that for a translation invariant system with post-quench Hamiltonian $H=H_0+\lambda\int d{\bf x}\,\Psi({\bf x})$, the large time limit of the one-point function of an operator $\Phi$ takes the form
\EQ
\langle\Phi({\bf x},t)\rangle=\langle\Phi\rangle_\lambda^\textrm{eq}+\lambda\left[\frac{2}{M^2}F_{1}^\Psi F_{1}^\Phi\,\cos Mt+O(t^{-\alpha})\right]+O(\lambda^2)\,,
\label{hom1}
\EN
where $\langle\Phi\rangle_\lambda^\textrm{eq}$ is the equilibrium expectation value in the theory with the Hamiltonian $H$, $M>0$ the quasiparticle mass, $F_{1}^{\cal O}$ the matrix element of ${\cal O}$ between $|0\rangle$ and the single-quasiparticle state, and 
\EQ
\alpha\geq d/2\,.
\label{alpha}
\EN
Throughout the paper we adopt natural units in which the maximal velocity of the quasiparticles is $v_\textrm{max}=1$.
In case of systems possessing several quasiparticle species the term in the square bracket is summed over the species. (\ref{hom1}) shows, in particular, the presence of undamped oscillations under the same conditions determined in \cite{quench} for the case $d=1$ (for which $\alpha$ is generically $3/2$), namely when the excitations produced by the quench include a single-quasiparticle mode\footnote{This in turn requires interacting quasiparticles, otherwise $\Psi$ creates only quasiparticle pairs.} ($F_{1}^\Psi\neq 0$) and the observable couples to this mode ($F_{1}^\Phi\neq 0$). 

In turn, (\ref{hom1}) is a particular case (${\cal D}=\mathbb{R}^d$) of the more general situation that we are going to consider, namely that of a quench performed only in a subregion ${\cal D}$ of the full space $\mathbb{R}^d$ occupied by the system. For such an inhomogeneous quench, corresponding to the post-quench Hamiltonian
\EQ
H=H_0+\lambda\int_{\cal D} d{\bf x}\,\Psi({\bf x})\,,
\label{H}
\EN
we show that the large time limit (\ref{hom1}) generalizes to
\bea
&& \langle\Phi({\bf x},t)\rangle= \langle\Phi({\bf x})\rangle_\lambda^\textrm{eq} \label{inhom1}\\
&& \hspace{.5cm} +\lambda\left[\frac{2F_{1}^\Psi F_{1}^\Phi}{(2\pi)^d}\int_{\cal D} d{\bf y}\int d{\bf p}\,\frac{\cos(\sqrt{{\bf p}^2+M^2}\,t+({\bf x}-{\bf y})\cdot {\bf p})}{{\bf p}^2+M^2}+O(t^{-\alpha-\beta_{\cal D}})\right]+O(\lambda^2),\nonumber
\eea
where $\alpha$ is the same entering (\ref{hom1}), while
\EQ
\beta_{\cal D}\in[0,d/2]
\label{beta}
\EN
rules the large time behavior $t^{-\beta_{\cal D}}$ of the integral term. If $L^d$, with $L\to \infty$, is the volume of $\mathbb{R}^d$, there are undamped oscillations ($\beta_{\cal D}=0$) when the volume of ${\cal D}$, vol(${\cal D}$), is of order $L^d$, a condition weaker than ${\cal D}=\mathbb{R}^d$. On the other hand, $\beta_{\cal D}=d/2$ when vol(${\cal D}$) is finite. Intermediate values of $\beta_{\cal D}$ occur depending on the number of dimensions in which vol(${\cal D}$) is extensive. We also show that the effect of the quench is appreciable only inside a light cone that at time $t$ contains ${\cal D}$ and the external region within distance $t$ from the boundary $\partial{\cal D}$. 

These analytical results -- whose degree of generality is uncommon in the framework of nonequilibrium quantum dynamics -- lead to the following physical deductions. The remainder of (\ref{hom1}) (terms of order $\lambda^2$ and higher) has one of the following behaviors for $t\to\infty$: indefinite growth in modulus, approach to a constant, or undamped oscillations. Hence, if we exclude that $\langle\Phi({\bf x},t)\rangle$ indefinitely grows in modulus -- something that is not expected on physical grounds -- (\ref{hom1}) leads to undamped oscillations also when $\lambda$ is not small (provided that $F_{1}^\Psi F_{1}^\Phi\neq 0$). The same applies to (\ref{inhom1}) when $\beta_{\cal D}=0$. To understand the role of ${\cal D}$ in this respect, we observe that the oscillations propagate inside the light cone, and are sustained by the energy produced in the quench and carried by the quasiparticles. In order the oscillations to stay undamped, a nonzero energy density inside the light cone is needed at late times (besides $F_{1}^\Psi F_{1}^\Phi\neq 0$). Since the energy produced by the quench and conserved by the time evolution is proportional to vol(${\cal D}$), and since the volume enclosed by the light cone becomes $L^d$ as $t\to\infty$, a nonzero energy density at late times requires that vol(${\cal D}$) is of order $L^d$, and this is the case corresponding to $\beta_{\cal D}=0$. 

The paper is organized as follows. The main theoretical results are derived in the next section, while in section~3 we illustrate the features of the time evolution also with a number of examples of quenched regions ${\cal D}$. Section~4 illustrates the role played by internal symmetries through the basic example of the quantum Ising model, and the last section contains some concluding remarks.

\section{Quenches in $d$ dimensions}
\subsection{Post-quench state and one-point functions}
We consider a $d$-dimensional system occupying the whole space $\mathbb{R}^d$. Before the quench the system is translation invariant and in the ground state $|0\rangle$ of the Hamiltonian $H_0$. We perform the theoretical analysis exploiting the complete basis of asymptotic quasiparticle states $|{\bf p}_1,\ldots,{\bf p}_n\rangle$ of the pre-quench theory, with ${\bf p}_i$ denoting the $d$-dimensional momenta of the quasiparticles. The asymptotic states are eigenstates of $H_0$ with eigenvalues equal to the sum of the quasiparticle energies $E_{{\bf p}_i}=\sqrt{M^2+{\bf p}_i^2}$. $M>0$ is the quasiparticle mass and measures the distance from a quantum critical point. In order to simplify the notation we refer to the case of a single quasiparticle species; generalizations are straightforward and will be discussed when relevant. 

The quench at $t=0$ is performed changing the Hamiltonian to (\ref{H}), and for ${\cal D}\neq\mathbb{R}^d$ breaks translation invariance. Since the quench excites quasiparticle modes, the pre-quench state $|0\rangle$ evolves into the state $|\psi_0\rangle=S_\lambda|0\rangle$, where 
\EQ
S_\lambda=T\,\exp\left(-i\lambda\int_0^\infty dt\int_{\cal D} d{\bf x}\,\Psi({\bf x},t)\right)
\label{Slambda}
\EN
($T$ denotes chronological ordering) is the operator whose matrix elements $\langle n|S_\lambda|0\rangle$ give the probability amplitude that the quench induces the transition from $|0\rangle$ to $|n\rangle$. Here we adopt the compact notation $|n\rangle=|{\bf p}_1,\ldots,{\bf p}_n\rangle$. To first order in the quench parameter $\lambda$ we have
\begin{equation}
|\psi_0\rangle\simeq |0\rangle + \lambda \SumInt\limits_{n,\mathbf{p}_i} \frac{g_{\mathcal{D}}(\mathbf{P})}{E} [F_n^{\Psi}]^* |n\rangle\,,
\label{psi0}
\end{equation}
where we defined 
\EQ
E=\sum_{i=1}^nE_{{\bf p}_i}\,,\hspace{1cm} {\bf P}=\sum_{i=1}^n{\bf p}_i\,,
\EN
\begin{equation}
g_{\mathcal{D}}(\mathbf{P})= \int_{\mathcal{D}}d\mathbf{x}\, e^{i \mathbf{P}\cdot \mathbf{x}}\,,
\label{g_D}
\end{equation}
\begin{equation}
F_n^{\cal O}(\mathbf{p}_1,...,\mathbf{p}_n)=\langle 0|{\cal O}(0,0)|\mathbf{p}_1,...,\mathbf{p}_n\rangle\,,
\label{ff}
\end{equation}
introduced the notation
\begin{equation}
\SumInt\limits_{n,\mathbf{p}_i}=\sum_{n=1}^{\infty} \frac{1}{n!} \int_{-\infty}^{\infty} \prod_{i=1}^n \frac{d\mathbf{p}_i}{(2\pi)^d E_{\mathbf{p}_i}}\,,
\label{sumint}
\end{equation}
and used
\begin{equation}
{\cal O}(\mathbf{x},t)=e^{i \mathcal{P}\cdot \mathbf{x}+iH_0 t}\,{\cal O}(0,0)\, e^{-i\mathcal{P}\cdot \mathbf{x}-iH_0 t}\,,
\label{translation}
\end{equation}
with $\mathcal{P}$ the momentum operator and ${\cal O}$ a generic local operator. As usual, an infinitesimal imaginary part is given to the energy to make the time integral in (\ref{Slambda}) convergent\footnote{The sum in (\ref{sumint}) starts from $n=1$ rather than from $n=0$ because the $O(\lambda)$ contribution to (\ref{psi0}) with $n=0$ (corresponding to $E={\bf P}=0$) diverges and must be subtracted. Such a term corresponds to vacuum energy renormalization and can be canceled through a counterterm in the Hamiltonian \cite{quench}.}. The result (\ref{psi0}) shows that the quench produces excitation modes with any number of quasiparticles and all possible momenta. Only when the quasiparticles do not interact, so that $H_0$ and $H$ are quadratic in the quasiparticle modes and $F_n^\Psi\propto\delta_{n,2}$, a post-quench state with quasiparticles organized in pairs is obtained.

The one-point function of a local observable $\Phi$ is given by the expectation value $\langle\Phi({\bf x},t)\rangle$ on the post-quench state (\ref{psi0}). In the formalism of asymptotic states the space-time dependence is carried by the operator and is extracted exploiting (\ref{translation}). The variation $\delta\langle\Phi({\bf x},t)\rangle$ of the one-point function of a hermitian observable with respect to the pre-quench value is given, at first order in $\lambda$, by 
\begin{align}
\delta\langle \Phi(\mathbf{x},t)\rangle &=\langle \psi_0|\Phi(\mathbf{x},t)|\psi_0\rangle-\langle 0|\Phi(0,0)|0\rangle +C_{\Phi}(\mathbf{x}) \nonumber\\
&\simeq 2\lambda \SumInt\limits_{n,\mathbf{p}_i} \frac{1}{E}\, \text{Re}\left\{ g_{\mathcal{D}}(\mathbf{P})\, [F_n^{\Psi}]^* F_n^{\Phi}\, e^{-i(Et+\mathbf{P}\cdot \mathbf{x})}\right\} +C_{\Phi}(\mathbf{x})\,,
\label{1point}
\end{align}
where we took into account that normalizing by $\langle\psi_0|\psi_0\rangle=1+{O}(\lambda^2)$ is immaterial at first order, and added the term
\begin{equation}
C_{\Phi}(\mathbf{x})\simeq -2\lambda \SumInt\limits_{n,\mathbf{p}_i} \frac{1}{E}\, \text{Re}\left\{ g_{\mathcal{D}}(\mathbf{P})\, [F_n^{\Psi}]^* F_n^{\Phi}\, e^{-i\mathbf{P}\cdot \mathbf{x}}\right\}
\label{C}
\end{equation}
to ensure continuity at $t=0$, namely the condition $\delta\langle\Phi({\bf x},0)\rangle=0$, which has no reason to be automatically satisfied.

\subsection{Light cone from the quenched domain and behavior at large times}
\label{largetime}
\begin{figure}[t]
    \centering
    \begin{subfigure}[h]{0.45\textwidth}
        \includegraphics[width=\textwidth]{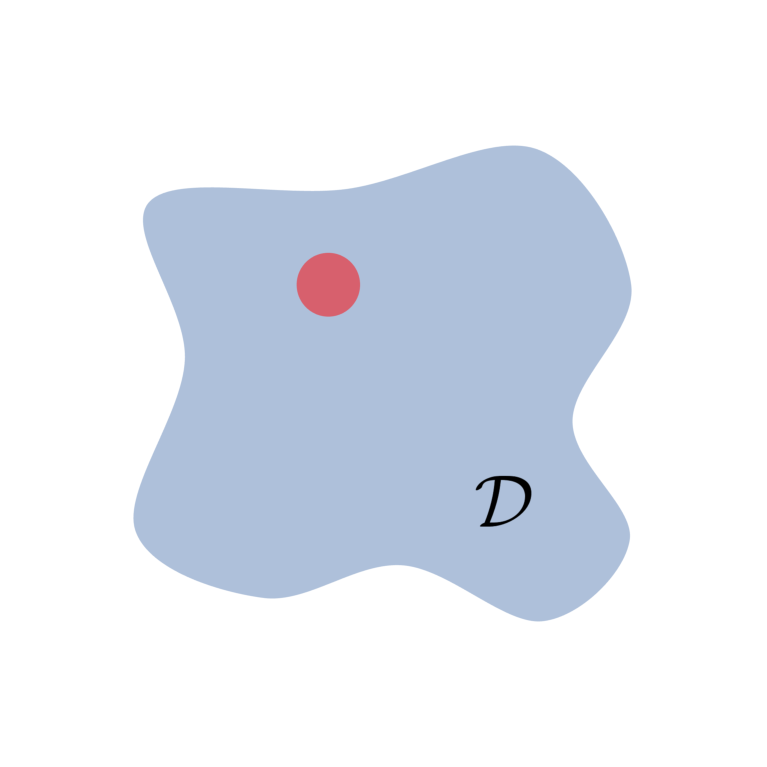}
    \end{subfigure}\hspace{1cm}%
    \begin{subfigure}[h]{0.45\textwidth}
        \includegraphics[width=\textwidth]{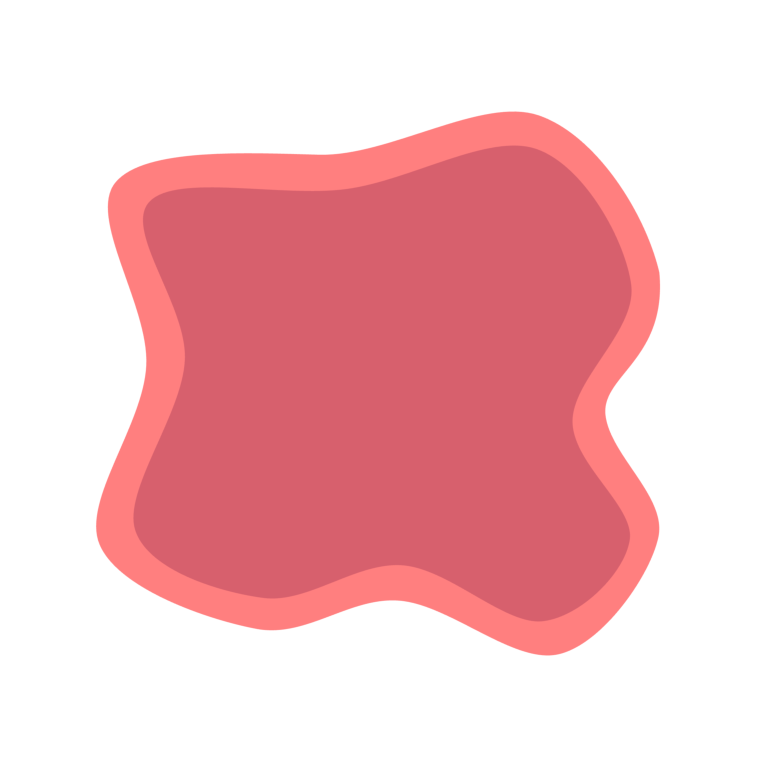}
    \end{subfigure}
    \caption{Implications of the stationarity condition (\ref{stationary}) for a quench in a region ${\cal D}$ of the space $\mathbb{R}^d$ occupied by the system. \textbf{Left:} The quench operator $\Psi$ acting at a point ${\bf y}\in{\cal D}$ creates excitations that at time $t$ have spread inside a sphere of radius $t$.  \textbf{Right:} Since $\Psi$ acts at all points of ${\cal D}$, the effect of the quench is appreciable only inside the light cone containing ${\cal D}$ and the region within distance $t$ from its boundary. }
    \label{balls}
\end{figure}

We can use (\ref{g_D}) to rewrite (\ref{1point}) as
\EQ
\delta\langle \Phi(\mathbf{x},t)\rangle \simeq 2\lambda \SumInt\limits_{n,\mathbf{p}_i} \int_{\mathcal{D}}d\mathbf{y}\, \frac{1}{E}\,\text{Re}\left\{ [F_n^{\Psi}]^* F_n^{\Phi}\, e^{-i[Et+\mathbf{P}\cdot (\mathbf{x}-\mathbf{y})]}\right\} +C_{\Phi}(\mathbf{x})\nonumber\,.
\label{1point1}
\EN
For $t$ large the rapid oscillation of the exponential suppresses the integrals over momenta unless the phase is stationary, namely unless 
\begin{equation}
\nabla_{\mathbf{p}_i} [E_{\mathbf{p}_i} t+\mathbf{p}_i\cdot (\mathbf{x}-\mathbf{y})]=\mathbf{v}_i t+\mathbf{x}-\mathbf{y}=0\,, \qquad i=1,2,\ldots,n\,,
\label{stationary}
\end{equation}
where we introduced the quasiparticle velocities \EQ\mathbf{v}_i=\nabla_{\mathbf{p}_i}E_{\mathbf{p}_i}=\frac{\mathbf{p}_i}{\sqrt{M^2+\mathbf{p}_i^2}}\,.
\EN
Since $|\mathbf{v}_i|<1$, the stationarity condition (\ref{stationary}) is satisfied when 
\begin{equation}
|\mathbf{x}-\mathbf{y}|<t\,.
\label{lightcone}
\end{equation}
This means that, for any point ${\bf y}\in{\cal D}$, the effect of the quench is appreciable only within a distance $t$ from ${\bf y}$, namely the maximal distance that the quasiparticles excited by the quench at the point ${\bf y}$ could reach at time $t$. Hence, the time evolution takes place inside a light cone\footnote{See \cite{lightcone} for the derivation of the light cone associated to the spreading of two-point correlations in the translation invariant case, which involves the connectedness properties of matrix elements as an additional ingredient.} containing ${\cal D}$ and the external region within distance $t$ from $\partial{\cal D}$ (figure~\ref{balls}).

For ${\cal D}$ finite, ${\bf x}$ fixed and $t$ large enough, the stationarity condition $\mathbf{v}_i=(\mathbf{y}-\mathbf{x})/t$ implies that the time dependence in (\ref{1point}) receives a significant contribution only when all momenta ${\bf p}_i$ are small, and then can be evaluated with $g_{\cal D}\to\textrm{vol}({\cal D})$ and $E_{{\bf p}_i}\to M+{\bf p}_i^2/2M$. The matrix elements (\ref{ff}) do not diverge at small momenta (see e.g. \cite{Barton}), and in such a limit $[F_n^\Psi]^*F_n^\Phi$ will behave as momentum to a power $2\alpha_n\geq 0$. In particular $\alpha_1=0$, since $F_1^{\cal O}$ is a real constant for the scalar hermitian operators that we consider. With this information, it is easy to rescale the momenta and see that the $n$-quasiparticle contribution in (\ref{1point}) behaves at large times as $t^{-(nd/2+\alpha_n)}$. 

More generally, suppose that ${\cal D}$ goes from $-\infty$ to $+\infty$ in $k$ of the $d$ spatial dimensions (say $x_1,x_2,\ldots,x_k$). Now $g_{\cal D}\propto\delta(P_1)\cdots\delta(P_k)$, and this gives an extra contribution $-k$ to the counting of the powers of momentum, so that the large time behavior is modified to $t^{-[(nd-k)/2+\alpha_n]}$. As long as $F_1^\Psi F_1^\Phi\neq 0$ the leading contribution comes from the single-quasiparticle mode $n=1$ and goes as $t^{-(d-k)/2}$. For an homogeneous quench (${\cal D}=\mathbb{R}^d$, i.e. $k=d$) we have undamped oscillations and the result in the square bracket of (\ref{hom1}), with the lower bound (\ref{alpha}) coming from $n=2$. A generic $k$ gives the time dependence in the square bracket of (\ref{inhom1}), behaving as $t^{-\beta_{\cal D}}$, with $\beta_{\cal D}=(d-k)/2$ that satisfies (\ref{beta}).  

On the other hand, suppose that ${\cal D}$ differs from $\mathbb{R}^d$ for the subtraction of a finite region, namely ${\cal D}\cup\tilde{\cal D}=\mathbb{R}^d$ with $\tilde{\cal D}$ finite. In this case $g_{\cal D}=g_{\mathbb{R}^d}-g_{\tilde{\cal D}}$, and the dependence at large times corresponds to the difference of the previous cases $k=d$ and $k=0$. Hence, we have $\beta_{\cal D}=0$ and undamped oscillations at sufficiently large times produced by the integral in (\ref{inhom1}). The conclusion that in general there are undamped oscillations when 
\EQ
\rho_{\cal D}=\textrm{vol}({\cal D})/\textrm{vol}(\mathbb{R}^d)
\label{rho}
\EN
is nonzero corresponds to the physical picture anticipated in the introduction: the energy produced by the quench is proportional to $\textrm{vol}({\cal D})$ and spreads in time inside the light cone, so that the energy density will be asymptotically proportional to $\rho_{\cal D}$; a nonzero $\rho_{\cal D}$ is able to keep the oscillations undamped. The amplitude of the undamped oscillations goes to zero if $\rho_{\cal D}$ goes to zero. The condition $\rho_{\cal D}>0$ amounts to $\textrm{vol}({\cal D})$ extensive in all dimensions.

We finally explain the origin of the time-independent term in (\ref{hom1}) and (\ref{inhom1}). For this purpose observe that in the {\it equilibrium} theory with Hamiltonian (\ref{H}) the first order contribution in $\lambda$ to the one-point function $\langle\Phi({\bf x})\rangle_\lambda^{\textrm{eq}}$ is\footnote{The subscript $c$ indicates the connected part of the two-point function.}
\begin{align}
\delta\langle \Phi(\mathbf{x})\rangle_{\lambda}^{\text{eq}} &\simeq -i\lambda \int_{-\infty}^{+\infty} dt \int_{\mathcal{D}} d\mathbf{y}\, \langle 0|T\,\Psi(\mathbf{y},t) \Phi(\mathbf{x},0)|0\rangle_c \nonumber\\
&= -2\lambda \SumInt\limits_{n,\mathbf{p}_i} \frac{1}{E}\, \text{Re}\left\{ [F_n^{\Psi}]^* F_n^{\Phi} \int_{\mathcal{D}} d\mathbf{y}\, e^{-i\mathbf{P}\cdot (\mathbf{x}-\mathbf{y})}\right\} \nonumber\\
&= C_{\Phi}(\mathbf{x})\,,
\label{deltaeq2}
\end{align}
where we used (\ref{translation}), expanded over asymptotic states, and finally compared with (\ref{C}). Hence, recalling what we just concluded about the time-dependent part, we have
\begin{equation}
\lim_{t\to\infty} \langle\Phi(\mathbf{x},t)\rangle=\langle \Phi(\mathbf{x})\rangle^{\text{eq}}_{\lambda}+O(\lambda^2)\,
\label{offset}
\end{equation}
when $\beta_{\cal D}\neq 0$. When $\beta_{\cal D}=0$, namely when $\textrm{vol}({\cal D})$ is extensive in all dimensions, the r.h.s. of (\ref{offset}) is the value around which the undamped oscillations take place.

\section{Following the time evolution}
\subsection{General features}
We rewrite (\ref{inhom1}) in the form
\begin{equation}
\langle \Phi(\mathbf{x},t)\rangle = \langle \Phi(\mathbf{x})\rangle_{\lambda}^{\text{eq}} + \lambda \left[\frac{2}{M^2}\,  F_1^{\Psi} F_1^{\Phi}\, f(\mathbf{x},t) +O(t^{-(\alpha+\beta_{\cal D})})\right]+O(\lambda^2)\,,
\label{inhom2}
\end{equation}
where
\EQ
f(\mathbf{x},t)= \frac{M^2}{(2\pi)^d} \int d\mathbf{p}\, \frac{1}{{\bf p}^2+M^2}\, \text{Re}\left\{ g_{\mathcal{D}}(\mathbf{p})\, e^{-i(\sqrt{\mathbf{p}^2+M^2}\,t+\mathbf{p}\cdot \mathbf{x})}\right\}
\label{1part}
\EN
is a dimensionless function. For $F_{1}^\Psi F_{1}^\Phi\neq 0$, $f({\bf x},t)$ determines the large time behavior of the one-point function for small quenches, until a time scale $t_\lambda$ that goes to infinity as $\lambda$ is reduced\footnote{The expression of $t_\lambda$ was given in \cite{quench} for $d=1$. Its generalization to the present $d$-dimensional case is $t_\lambda\sim 1/\lambda^{1/(d+1-X_\Psi)}$, where $X_\Psi<d+1$ is the scaling dimension of $\Psi$ at the quantum critical point. \label{scale}}. On the other hand, the analysis of section \ref{largetime} does not allow to neglect the terms with $n>1$ in (\ref{1point}) when $t$ is {\it not} large. However, it is known that in $d=1$ \cite{DV,oscill} the contribution of the $n$-quasiparticle state is normally rapidly suppressed as $n$ increases, so that $f({\bf x},t)$ provides a good approximation also for short times. If this is true also for $d>1$ we should expect, recalling (\ref{deltaeq2}) and (\ref{C}), that the first order contribution to the equilibrium expectation value is well approximated as
\EQ
\delta\langle\Phi({\bf x})\rangle_\lambda^{\textrm{eq}}\approx -\lambda\,\frac{2}{M^2}\,F_{1}^\Psi F_{1}^\Phi\,f({\bf x},0)\,.
\label{approx}
\EN
We will soon explicitly illustrate that (\ref{approx}) indeed yields the result expected for $\delta\langle\Phi({\bf x})\rangle_\lambda^{\textrm{eq}}$, namely a function that is essentially constant for ${\bf x}\in{\cal D}$ and zero otherwise. With this anticipation, we see that (\ref{inhom2}) (without the term $O(t^{-\alpha-\beta_{\cal D}})$) can be used not only for large times but, with good approximation, also for short times, meaning that the function (\ref{1part}) yields a global view of the time evolution for small quenches. We now consider this function for a number of quenching domains ${\cal D}$.

\begin{figure}[t]
    \centering
    \begin{subfigure}[h]{0.33\textwidth}
        \includegraphics[width=\textwidth]{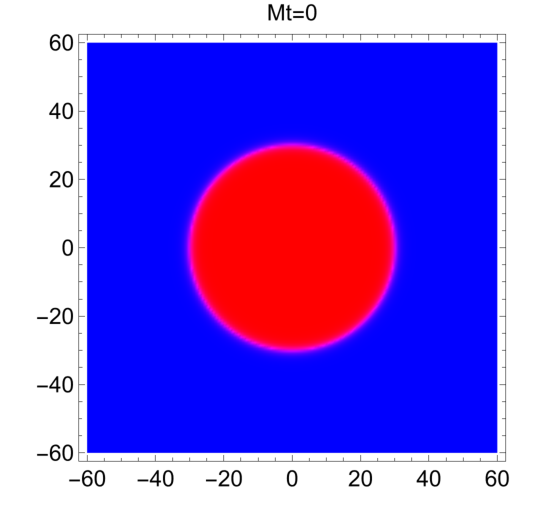}
    \end{subfigure}\hspace{0.01cm}%
    \begin{subfigure}[h]{0.33\textwidth}
        \includegraphics[width=\textwidth]{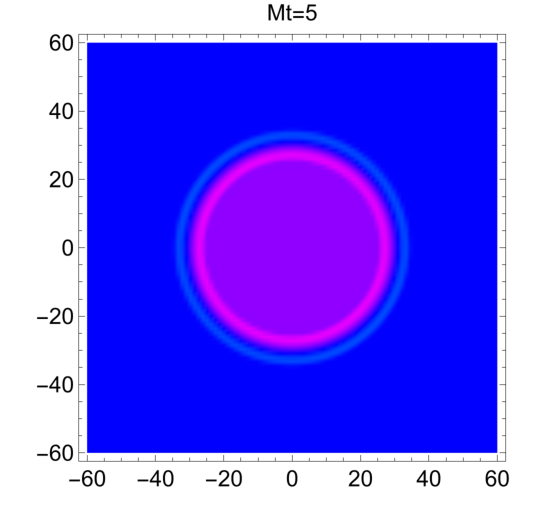}
    \end{subfigure}\hspace{0.01cm}%
    \begin{subfigure}[h]{0.33\textwidth}
        \includegraphics[width=\textwidth]{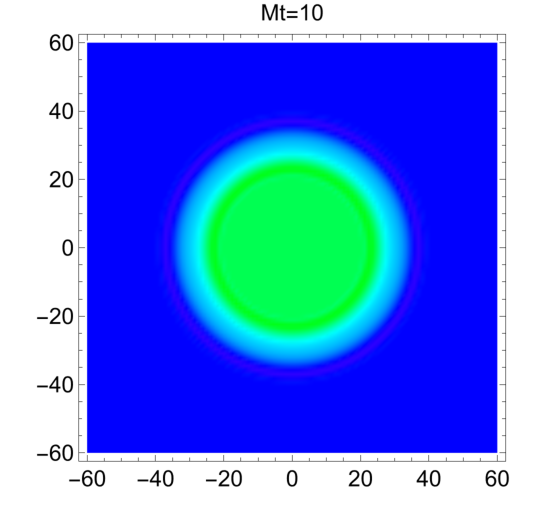}
    \end{subfigure}
    \begin{subfigure}[h]{0.33\textwidth}
        \includegraphics[width=\textwidth]{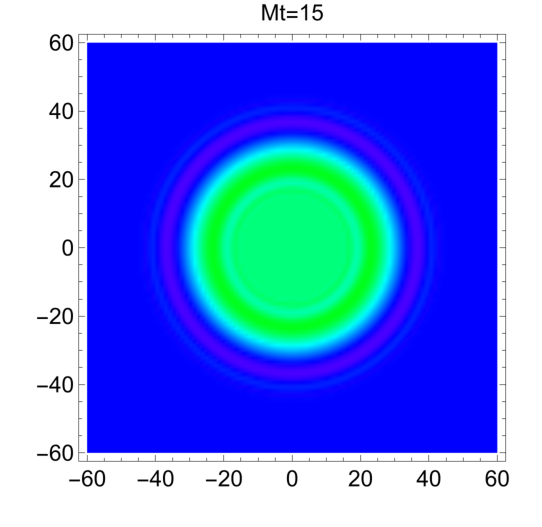}
    \end{subfigure}\hspace{0.01cm}%
    \begin{subfigure}[h]{0.33\textwidth}
        \includegraphics[width=\textwidth]{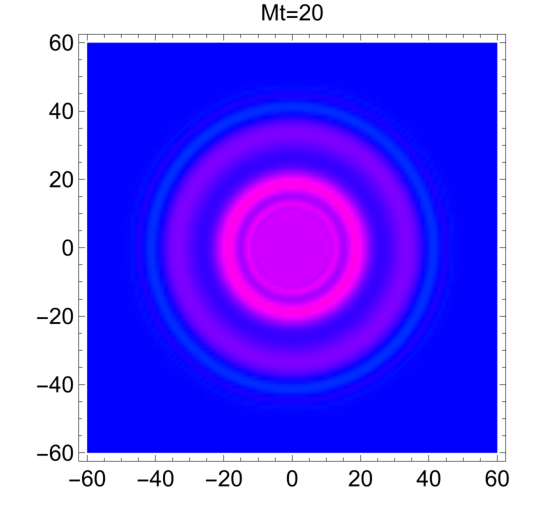}
    \end{subfigure}\hspace{0.01cm}%
    \begin{subfigure}[h]{0.33\textwidth}
        \includegraphics[width=\textwidth]{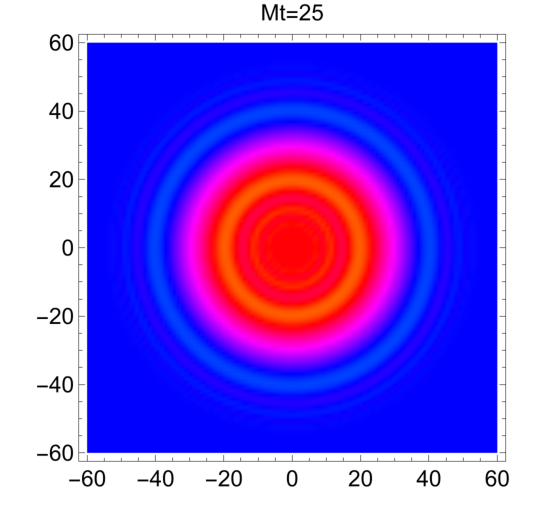}
    \end{subfigure}
    \begin{subfigure}[h]{0.33\textwidth}
        \includegraphics[width=\textwidth]{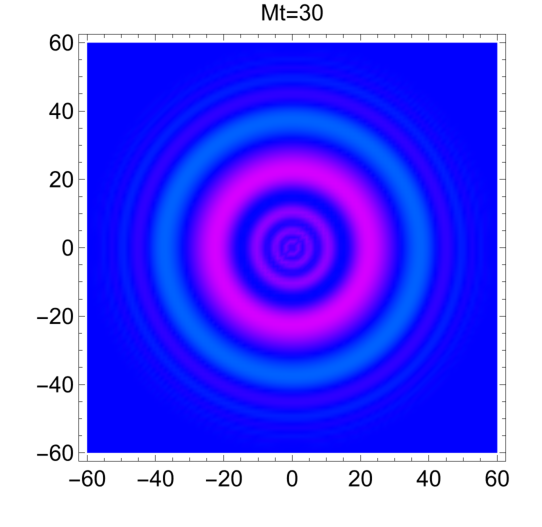}
    \end{subfigure}\hspace{0.01cm}%
    \begin{subfigure}[h]{0.33\textwidth}
        \includegraphics[width=\textwidth]{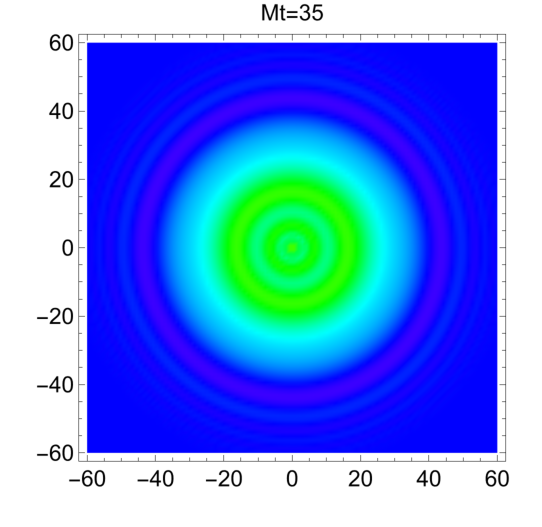}
    \end{subfigure}\hspace{0.01cm}%
    \begin{subfigure}[h]{0.33\textwidth}
        \includegraphics[width=\textwidth]{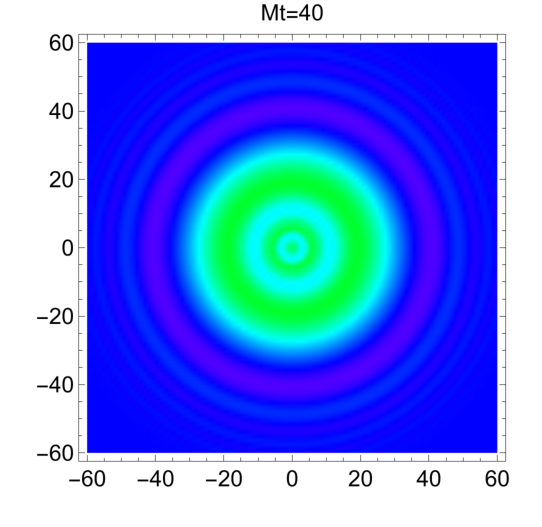}
    \end{subfigure}
     \begin{subfigure}[h]{0.23\textwidth}
        \includegraphics[width=\textwidth]{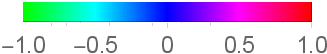}
    \end{subfigure}
    \caption{$f(\mathbf{x},t)$ at different times in the plane $Mx_1$-$Mx_2$ for a quench in $d=2$ with ${\cal D}$ a disk of radius $b=30/M$.}
    \label{disk_t}
\end{figure}

\begin{figure}[t]
    \centering
    \begin{subfigure}[h]{0.45\textwidth}
        \includegraphics[width=\textwidth]{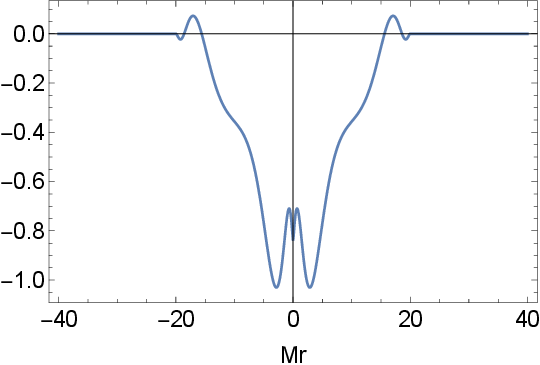}
    \end{subfigure}\hspace{1cm}%
    \begin{subfigure}[h]{0.45\textwidth}
        \includegraphics[width=\textwidth]{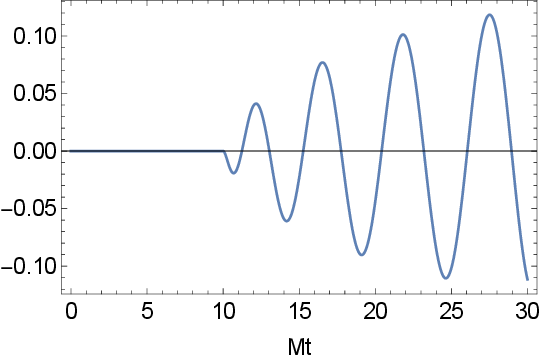}
    \end{subfigure}
    \caption{$f(\mathbf{x},t)$ for a quench in $d=2$ with ${\cal D}$ a disk of radius $b=10/M$.
    \textbf{Left:} $Mt=10$. The function is essentially zero outside the edges of the light cone located at distance $r=b+t$ from the origin.  \textbf{Right:} $Mr=20$.  Time evolution becomes appreciable only after that the light cone is reached at time $t= r-b$.}
    \label{disk_cone}
\vspace{1cm}
    \centering
    \begin{subfigure}[h]{0.5\textwidth}
        \includegraphics[width=\textwidth]{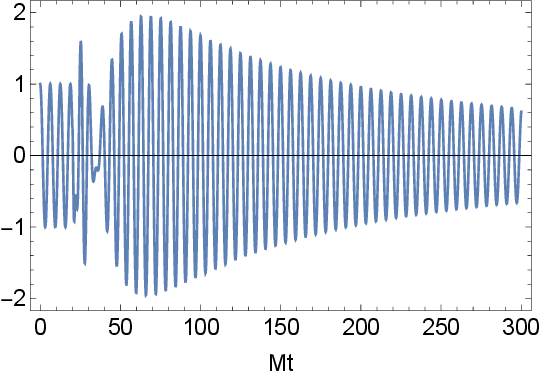}
    \end{subfigure}
    \caption{$f(0,t)$ for a quench in $d=2$ with ${\cal D}$ a disk of radius $b=20/M$. The oscillations in the origin stay undamped until the arrival at time $t=b$ of the modes originating from $\partial{\cal D}$. Eventually, the damping factor $t^{-d/2}$ prescribed by (\ref{inhom1}) with ${\cal D}$ finite sets in for large times.}
    \label{suppression}
\end{figure}

\subsection{Rotationally invariant quenched domains}
If ${\cal D}$ is the $d$-dimensional sphere of radius $b$ centered in the origin, (\ref{g_D}) and (\ref{1part}) yield
\begin{equation}
g_\textrm{sphere}(\mathbf{P})= \left(\frac{2\pi b}{|{\bf P}|} \right)^{\frac{d}{2}} J_{\frac{d}{2}}(|{\bf P}|b)\,,
\label{g_sphere}
\end{equation}
and
\EQ
f(\mathbf{x},t)= \frac{M^2\, b^{\frac{d}{2}}}{r^{\frac{d}{2}-1}} \int_0^{\infty} dp\,  J_{\frac{d}{2}}(pb) J_{\frac{d}{2}-1}(pr)\, \frac{\cos(\sqrt{M^2+p^2}\, t)}{M^2+p^2}\,,
\label{f_sphere}
\EN
respectively, with $J_\alpha(z)$ the Bessel function and $r=|{\bf x}|$. The function (\ref{f_sphere}) is plotted at different times in $d=2$ (${\cal D}=\textrm{disk}$) in figure~\ref{disk_t}. The result for $t=0$ confirms what we anticipated about (\ref{approx}) and the ability of $f({\bf x},t)$ to give an accurate view of the time evolution also for short times. As $t$ increases the figure clearly shows the spreading of the light cone located at distance $t$ from the boundary of the disk. The boundary modes propagate also inside the disk, but for $t<b$ they leave unaffected the central region with $r<b-t$. Here the function essentially behaves as for a homogeneous quench, namely is spatially constant with undamped oscillations in time. The presence of the light cone is also illustrated in figure~\ref{disk_cone}. Figure~\ref{suppression} shows the damping of the oscillations at large times, with the suppression $t^{-d/2}$ expected for the case ${\cal D}$ finite. 

\begin{figure}[t]
    \centering
    \begin{subfigure}[h]{0.4\textwidth}
        \includegraphics[width=\textwidth]{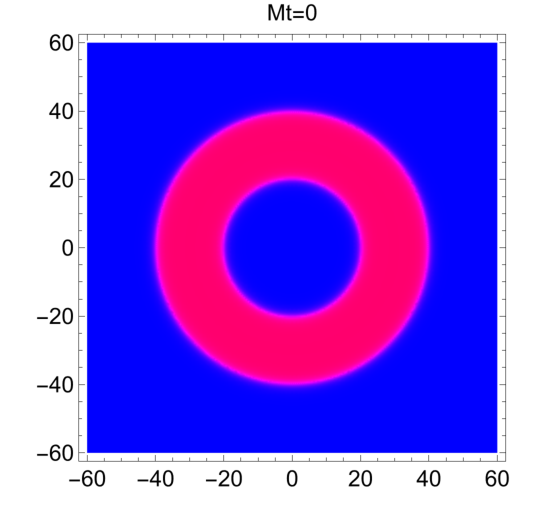}
    \end{subfigure}\hspace{0.01cm}%
    \begin{subfigure}[h]{0.4\textwidth}
        \includegraphics[width=\textwidth]{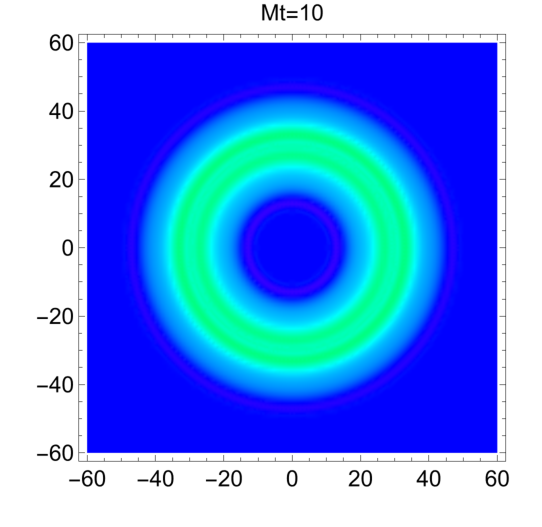}
    \end{subfigure}
    \begin{subfigure}[h]{0.4\textwidth}
        \includegraphics[width=\textwidth]{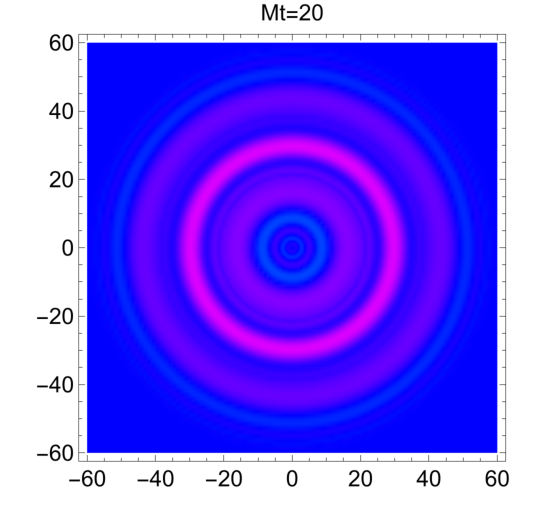}
    \end{subfigure}\hspace{0.01cm}%
    \begin{subfigure}[h]{0.4\textwidth}
        \includegraphics[width=\textwidth]{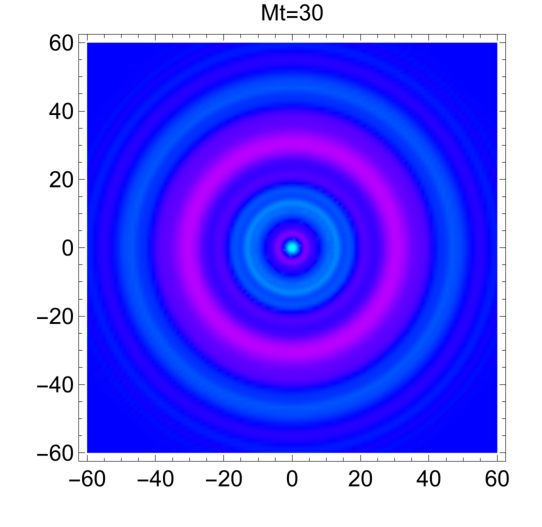}
    \end{subfigure}
     \begin{subfigure}[h]{0.25\textwidth}
        \includegraphics[width=\textwidth]{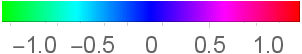}
    \end{subfigure}
    \caption{$f(\mathbf{x},t)$ at different times in the plane $Mx_1$-$Mx_2$ for a quench in $d=2$ with ${\cal D}$ an annulus occupying the region $20<Mr<40$. The initial central gap in the light cone closes when the boundary modes reach $r=0$ at $Mt=Mb_1=20$. See also figure~\ref{shell2}.}
    \label{shell1}
\end{figure}

\begin{figure}[t]
    \centering
        \includegraphics[width=9cm]{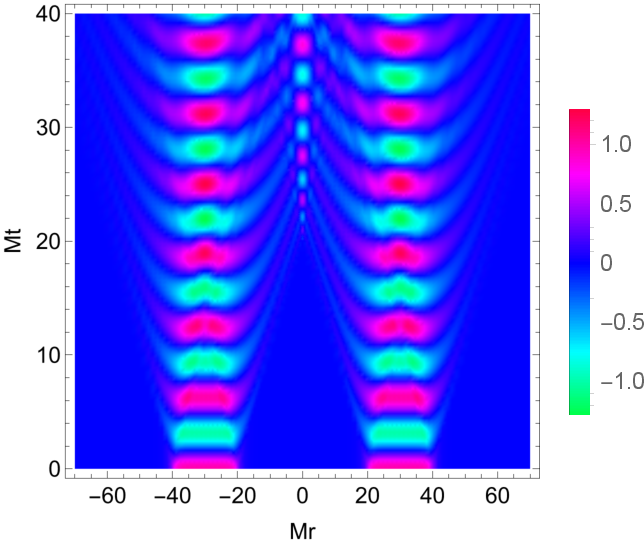}
    \caption{$f(\mathbf{x},t)$ for the same quench of figure~\ref{shell1}.}
    \label{shell2}
\end{figure}

\begin{figure}[h]
    \centering
        \includegraphics[width=9cm]{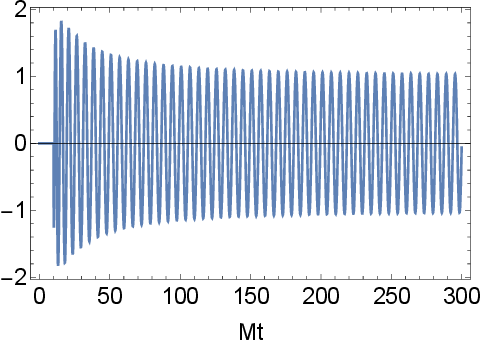}
    \caption{$f(0,t)$ for a quench in $d=3$ with ${\cal D}$ extending everywhere except a sphere of radius $b=10/M$ centered in the origin. The time evolution in the origin starts at $t=b$, with oscillations that become undamped at large time, as for the general case of vol$({\cal D})$ extensive in all dimensions.}
    \label{cavity}
\end{figure}

A straightforward generalization is that of a quench with ${\cal D}$ the $d$-dimensional spherical shell $b_1<r<b_2$, which yields
\begin{equation}
f(\mathbf{x},t)=\frac{M^2}{r^{\frac{d}{2}-1}} \int_0^{\infty} dp\,  \left[b_2^{\frac{d}{2}}\, J_{\frac{d}{2}}(pb_2)- b_1^{\frac{d}{2}}\, J_{\frac{d}{2}}(pb_1)\right] J_{\frac{d}{2}-1}(pr)\, \frac{\cos(\sqrt{p^2+M^2}\,t)}{p^2+M^2}\,.
\label{f_shell}
\end{equation}
The time evolution is illustrated in figures~\ref{shell1} and \ref{shell2} for $b_2$ finite, and in figure~\ref{cavity} for $b_2$ infinite. 

\subsection{Quenched domains with corners}
\begin{figure}[t]
    \centering
    \begin{subfigure}[h]{0.33\textwidth}
        \includegraphics[width=\textwidth]{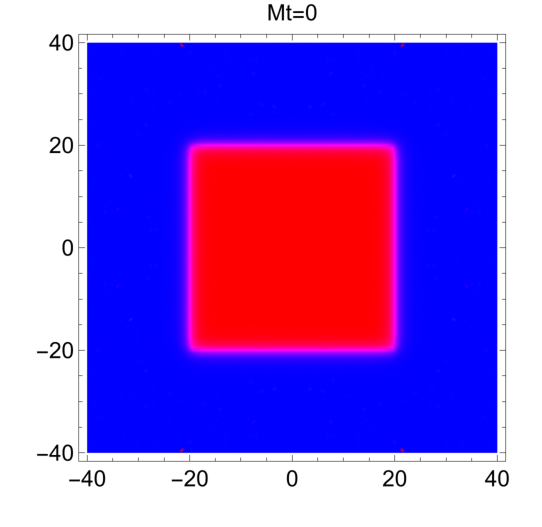}
    \end{subfigure}\hspace{0.01cm}%
    \begin{subfigure}[h]{0.33\textwidth}
        \includegraphics[width=\textwidth]{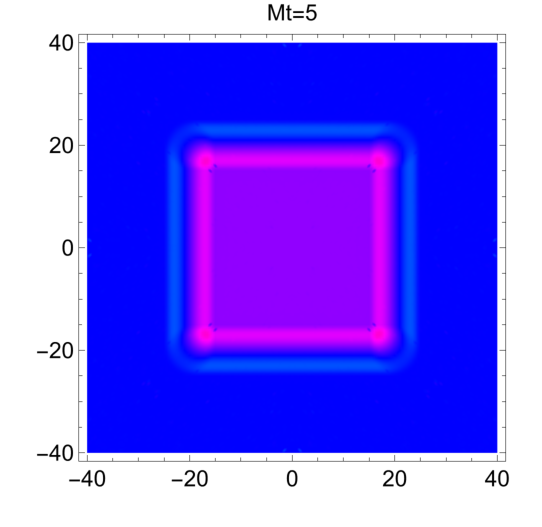}
    \end{subfigure}\hspace{0.01cm}%
    \begin{subfigure}[h]{0.33\textwidth}
        \includegraphics[width=\textwidth]{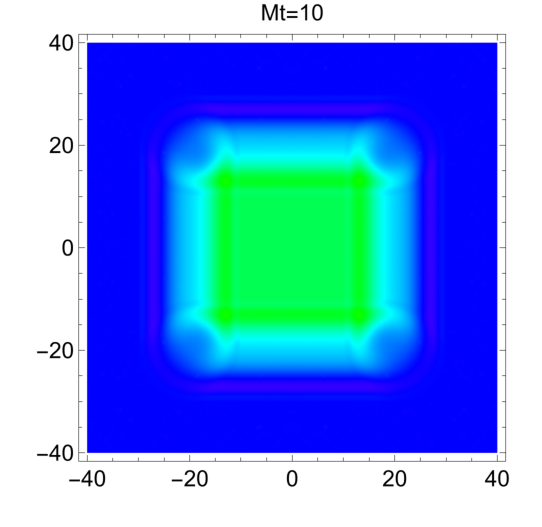}
    \end{subfigure}
    \begin{subfigure}[h]{0.33\textwidth}
        \includegraphics[width=\textwidth]{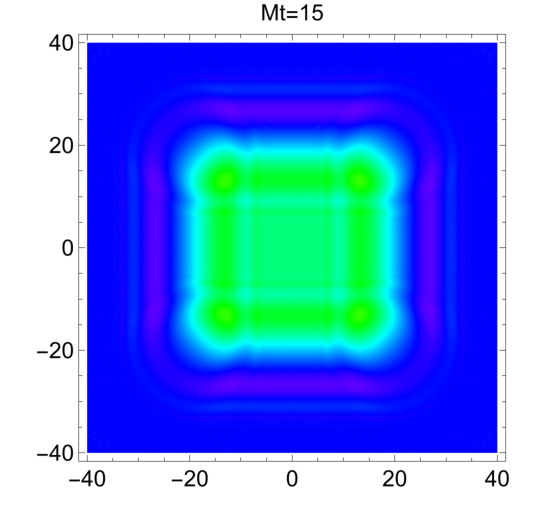}
    \end{subfigure}\hspace{0.01cm}%
    \begin{subfigure}[h]{0.33\textwidth}
        \includegraphics[width=\textwidth]{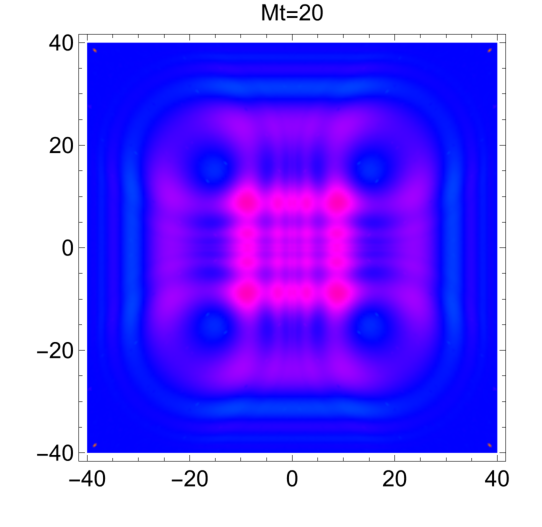}
    \end{subfigure}\hspace{0.01cm}%
    \begin{subfigure}[h]{0.33\textwidth}
        \includegraphics[width=\textwidth]{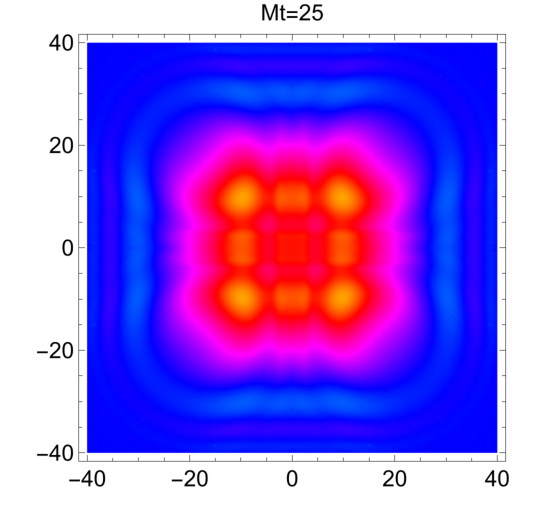}
    \end{subfigure}
    \begin{subfigure}[h]{0.33\textwidth}
        \includegraphics[width=\textwidth]{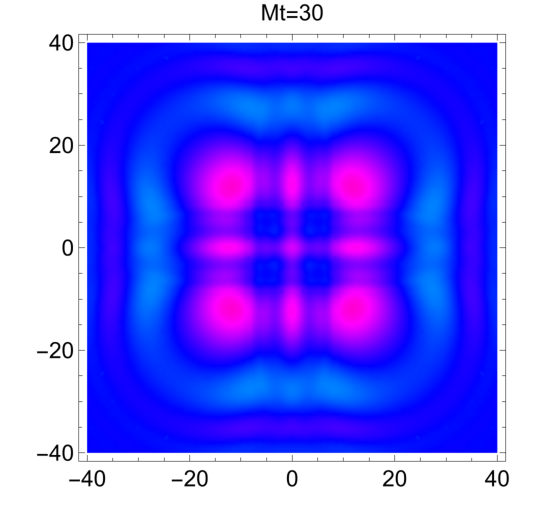}
    \end{subfigure}\hspace{0.01cm}%
    \begin{subfigure}[h]{0.33\textwidth}
        \includegraphics[width=\textwidth]{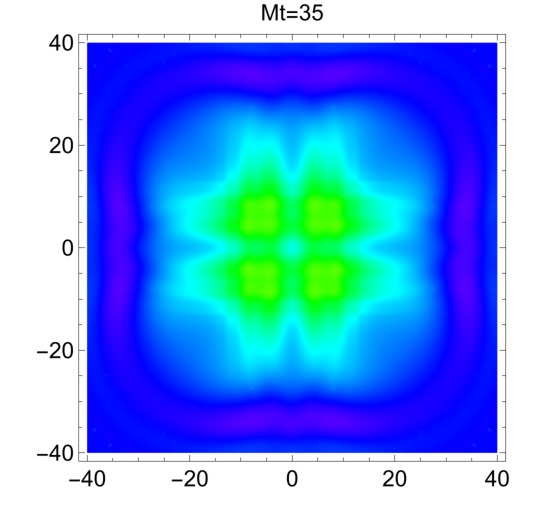}
    \end{subfigure}\hspace{0.01cm}%
    \begin{subfigure}[h]{0.33\textwidth}
        \includegraphics[width=\textwidth]{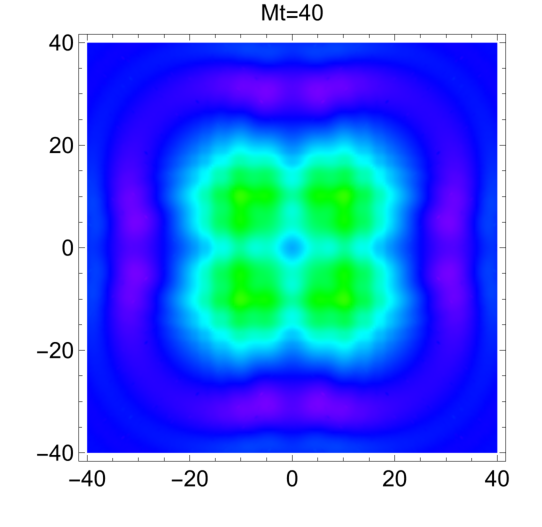}
    \end{subfigure}
     \begin{subfigure}[h]{0.23\textwidth}
        \includegraphics[width=\textwidth]{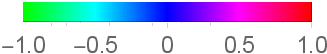}
    \end{subfigure}
    \caption{$f(\mathbf{x},t)$ at different times in the plane $Mx_1$-$Mx_2$ for a quench in $d=2$ with ${\cal D}$ a square of side $40/M$.}
    \label{square_t}
\end{figure}

If the quenched domain is a $d$-dimensional box, ${\cal D}=[-b_1,b_1]\times\cdots\times [-b_d,b_d]$, we have
\begin{equation}
g_\textrm{box}(\mathbf{P})=\prod_{k=1}^d \frac{2}{P_k}\,\sin(P_k b_k)\,.
\end{equation}
Figure~\ref{square_t} shows the corresponding function $f({\bf x},t)$ at different times in $d=2$ for a square domain. Again, the image at $t=0$, with the function essentially constant inside the square and vanishing outside, illustrates (through (\ref{approx})) that the time evolution is described with good approximation also at short times. For $t>0$, the general result that the light cone is located at distance $t$ from the boundary of ${\cal D}$ leads to a rounding in correspondence of the corners of the square. This example gives an idea of the patterns that can be expected when $\partial{\cal D}$ increasingly deviates from a smooth surface. 

\section{Role of internal symmetries: Ising model}
The model-dependent information for the use of (\ref{hom1}) and (\ref{inhom1}) is whether the matrix elements $F_1^\Psi$ and $F_1^{\Phi}$ vanish or not. They are generically nonzero, unless a symmetry forces them to vanish. In this section we illustrate the role of symmetries through the basic example of the $d$-dimensional quantum Ising ferromagnet. This is defined by the Hamiltonian
\EQ
H_\textrm{Ising}=-J\sum_{\langle i,j\rangle} \sigma^x_i\sigma^x_j-h_z\sum_i\sigma^z_i-h_x\sum_i\sigma^x_i\,,
\label{Ising}
\EN
where $\sigma^{x,y,z}_i$ are Pauli matrices at site $i$, $\langle i,j\rangle$ denotes a pair of nearest-neighbor sites, $J$ is positive, and $h_z$ and $h_x$ are the transverse and longitudinal magnetic fields, respectively. For $h_x=0$ and $|h_z|=h_z^c$ the system possesses a quantum critical point associated to the spontaneous breaking of spin reversal ($\mathbb{Z}_2$) symmetry in the $x$ direction and belonging to the universality class of the classical Ising model in $d+1$ dimensions. The operator $\sigma^z_i$ ($\sigma^x_i$) is $\mathbb{Z}_2$-even (odd). The paramagnetic (ferromagnetic) phase corresponds to $h_x=0$ and $|h_z|>h_z^c$ ($|h_z|<h_z^c$). 

\begin{figure}[t]
    \centering
        \includegraphics[width=10cm]{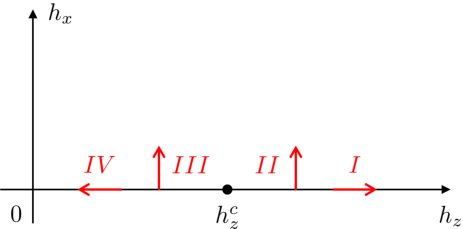}
    \caption{The different quenches in the quantum Ising model considered in the text. The critical point is located at $(h^c_z,0)$. With reference to the Hamiltonian (\ref{Ising}), each arrows goes from the pre-quench to the post-quench values of the parameters.}
    \label{Ising_quenches}
\end{figure}

We will consider quenches in which the system, which for $t<0$ is in the ground state $|0\rangle$ of the Hamiltonian $H_0$ given by (\ref{Ising}) with $h_x=0$, evolves for $t>0$ with the Hamiltonian (\ref{H}) with\footnote{Our theory is formulated in the continuum, which in the present case is accessed working not too far from the critical point.}  $\lambda\,\Psi({\bf x})$ equal to $\delta h_z\,\sigma^z({\bf x})$ ($\delta h_z\ll h_z$) or to $h_x\,\sigma^x({\bf x})$.  Depending on the fact that we start from the paramagnetic or ferromagnetic phase, we have the four quenches depicted in figure~\ref{Ising_quenches}. 

The specialization of (\ref{hom1}) and (\ref{inhom1}) to the different quenches proceeds through symmetry considerations in the pre-quench theory. In the paramagnetic phase, the fundamental quasiparticle excitation is created by the order parameter operator $\sigma^x({\bf x})$ and then is $\mathbb{Z}_2$-odd; it follows that $F_1^{\sigma^z}=0$ and $F_1^{\sigma^x}\neq 0$. In the ferromagnetic phase the symmetry is spontaneously broken and both $F_1^{\sigma^z}$ and $F_1^{\sigma^x}$ are nonzero in $d>1$. The case $d=1$ is special because the excitations in the ferromagnetic phase have a topological nature (they are kinks, see \cite{review} for a review) and can couple to $\sigma^x$ and $\sigma^z$ only in topologically neutral pairs, with the consequence that $F_1^{\sigma^z}=F_1^{\sigma^x}=0$. 

This information about $F_1^{\sigma^z}$ and $F_1^{\sigma^x}$ determines the time evolution of $\langle\sigma^x\rangle$ and $\langle\sigma^z\rangle$ through (\ref{hom1}) when the quenches\footnote{For quench I the symmetry implies $\langle\sigma^x\rangle=0$ at all orders in $\lambda$.} I-IV are performed in the whole space $\mathbb{R}^d$, and through (\ref{inhom1}) when they are performed only in a subregion ${\cal D}$. In particular we saw that, if vol$({\cal D})$ is extensive in all dimensions, undamped oscillations at large time occur when $F_1^\Psi F_1^{\Phi}\neq 0$, and we explicitly provide this information in table~\ref{Ising_table} for the case $\Phi=\sigma^x$. The peculiarity of the cases III and IV in $d=1$ follows from the previous observation about kinks. 

\begin{table}
\begin{center}
\begin{tabular}{|c||c|c|r|}
\hline
Quench & $\lambda$ & $\Psi$ & $F_1^\Psi F_1^{\sigma^x}\hspace{.7cm}$ \\
\hline
& & & $d=1$\hspace{1cm}$d>1$ \\
\hline\hline
{\color{red}{I}} & $\delta h_z$ & $\sigma^z$ & $0$\hspace{1.9cm}$0$ \\
\hline
{\color{red}{II}} &  $h_x$ & $\sigma^x$ & $\neq 0$\hspace{1.5cm}$\neq 0$ \\
\hline
{\color{red}{III}} & $h_x$ & $\sigma^x$ & $0$\hspace{1.5cm}$\neq 0$ \\
\hline
{\color{red}{IV}} & $\delta h_z$ & $\sigma^z$ & $0$\hspace{1.5cm}$\neq 0$ \\
\hline
\end{tabular}
\caption{Quenches in the $d$-dimensional quantum Ising ferromagnet indicated in figure~\ref{Ising_quenches}. The information about $F_1^\Psi F_1^{\sigma^x}$ determines the time evolution of the order parameter $\langle\sigma^x\rangle$ for homogeneous quenches through (\ref{hom1}), and for quenches in the subregion ${\cal D}$ through (\ref{inhom1}). The peculiarities of the case $d=1$ are discussed in the text.}
\label{Ising_table}
\end{center}
\end{table}

The topological nature of the excitations of the ferromagnetic phase in $d=1$ gives rise to an additional caveat about quench III in the Ising chain. While (\ref{hom1}) and (\ref{inhom1}) hold also in this case and do not yield undamped oscillations at first order in $\lambda=h_x$, the longitudinal field makes kinks unstable and confines them into topologically neutral pairs \cite{McW,DMS,review}, thus generating nonperturbatively the single-quasiparticle modes able to produce undamped oscillations of the order parameter on a time scale that becomes accessible for $h_x$ not too small\footnote{As recalled above, the results at first order in $\lambda$ quantitatively hold until a time scale that goes to infinity as $\lambda$ goes to zero. Ref.~\cite{DV} contains a more detailed discussion of Ising quenches in $d=1$, including the agreement with other analytical results \cite{CEF,ES} available for quenches I and IV (noninteracting fermions) when ${\cal D}=\mathbb{R}$. More generally, for $d=1$ one can exploit the exact knowledge of all matrix elements (\ref{ff}) when the pre-quench theory is integrable \cite{quench}. Following the suggestion of an anonymous referee, we also mention that it has been argued in \cite{CS} that the dynamics following a specific initial condition with exponential structure -- not originated by a quench -- can lead to oscillation damping.}. These oscillations have indeed been observed numerically in \cite{KCTC}. Hence, we see that for quench III kink confinement in $d=1$ produces for $h_x$ large enough the undamped oscillations that in $d>1$ are already present at first order in $h_x$. For quench IV, instead, the stability of the kinks precludes\footnote{Another manifestation of the different nature of the excitations in the spontaneously broken phase of the Ising model in $d=1$ and $d>1$ arises in the physics of interfaces \cite{ising3d,wall3d}.} in $d=1$ the undamped oscillations that arise in $d>1$. It is worth recalling that the presence of kinks in the spontaneously broken phase of systems with discrete symmetry, as well as their confinement under explicit symmetry breaking, are generic in $d=1$ (see \cite{Z3conf,LTD}), so that considerations analogous to those we made for Ising apply more generally. 

In $d=1$ the theory (\ref{Ising}) possesses a single species of quasiparticles at $h_x=0$, so that (\ref{hom1}) and (\ref{inhom1}) hold as they are. In $d>1$, on the other hand, more species may be present, in which case the term $2M^{-2}F_{1}^\Psi F_{1}^\Phi\cos Mt$ in (\ref{hom1}) is replaced by $\sum_a 2M_a^{-2}F_{1,a}^\Psi F_{1,a}^\Phi\cos M_at$, where $a$ labels the different species; a similar generalization occurs in (\ref{inhom1}). For the ferromagnetic phase in $d=2$ there is numerical consensus \cite{CH,LSW,CHPZ,DKSTV,Nishiyama,RBLD} about the existence, besides the lightest quasiparticle with mass $M_1$, of a second stable quasiparticle with mass $M_2\approx 1.8 M_1$. A spectral analysis of the undamped oscillations expected for quench IV should provide an alternative, possibly more accurate, way of determining this mass ratio. 

It was also observed in \cite{oscill} that in presence of several quasiparticle species oscillation frequencies $M_a-M_b$ can arise at order $\lambda^2$. In addition, the comparison performed in \cite{DV} with the result of \cite{ES} for noninteracting fermions indicates that corrections of order $\lambda^2$ and higher lead to the replacement of the pre-quench masses with the post-quench ones, and numerical evidence in this sense for interacting quasiparticles was given in \cite{Jacopo,HKT}. By post-quench masses we mean the masses of the {\it equilibrium} theory with the post-quench values of the couplings; their difference  from the pre-quench masses is of order $\lambda$. The mass ratio $M_2/M_1$ to be observed in quench IV in $d=2$ does not depend on $\lambda$ in the scaling region. 

The time evolution of the order parameter $\langle\sigma^x\rangle$ (in our notations) following a homogeneous quench in the Ising ferromagnetic phase has been numerically investigated in $d=2$ in \cite{HMH,HHM}, for a pre-quench value $h_z^i=0$ that maximizes the distance from the continuum limit of our analytical study. Still, confirming a robustness of our results, the plots show undamped oscillations, at least for post-quench values $h_z^f$ not too close to $h_z^c$. When $h_z^f$ approaches $h_z^c$ the correlation length becomes large, the long cylinder with a six-site circumference used in the simulation no longer approximates the plane, and a crossover to one-dimensional behavior (with damping of the oscillations) can be expected\footnote{We thank J.C. Halimeh for pointing out Refs. \cite{HMH,HHM} following the appearance of the preprint of this paper.}. In perspective, it would be very interesting to have numerical results for small quenches to perform the first quantitative comparison with analytical results in $d=2$.

\section{Conclusion}
We studied quantum quenches of systems in $d$ spatial dimensions that are initially in the ground state of a spatially homogeneous Hamiltonian $H_0$. The quench is performed instantaneously changing an interaction parameter inside a spatial region ${\cal D}$. The analytical results that we derived provide unique benchmarking for numerical and experimental methods, as well as a general picture of the time evolution of local observables. In the first place the evolution takes place inside a light cone that originates from the boundary of ${\cal D}$ at the moment of the quench and spreads outwards as time increases. Inside the light cone the observable undergoes oscillations with frequency equal to the quasiparticle mass, which persist undamped at late times under two types of conditions. The first type, of dynamical nature, requires that the state produced by the quench includes single-quasiparticle modes and that the observable couples to them. These requirements are generically fulfilled when the quasiparticles interact, unless internal symmetries of the system cause the vanishing of some matrix elements, a mechanism that we illustrated through the paradigmatic example of the quantum Ising model. The second condition, of geometrical type, is that the energy density does not tend to zero at large times, and is fulfilled when the volume of ${\cal D}$ is extensive in all dimensions. The wavefronts spreading from the boundary of ${\cal D}$ are increasingly structured as the boundary deviates from a smooth surface. Our formulae apply to any choice of ${\cal D}$, and we provided explicit illustrations of these features through some examples.

\end{document}